# The Double-edged Effect of Banning Generative AI on Online Question-and-Answer Communities: Evidence from Stack Exchange

Yuanhong Ma[1]*, Qinglai He[2]*, Xitong Li[3], Lynn Wu[4]

## Abstract

We investigate how banning generative artificial intelligence-generated content (AIGC) affects knowledge seeking, knowledge contribution, and contribution efficiency in online question-and-answer communities. After the launch of ChatGPT in late November 2022, several Stack Exchange communities implemented official bans on AIGC over concerns such as less reliable and socially engaged content. Leveraging data from the full network of Stack Exchange communities, we employ a difference-in-differences (DID) approach to examine the impacts of these bans. Our results reveal a double-edged impact: while the AIGC ban increases knowledge seeking, as evidenced by a higher volume of posted questions, it simultaneously reduces contribution efficiency, reflected in a lower proportion of questions receiving satisfactory answers within the expected time frame. Notably, these impacts are only evident in non-STEM communities. We take a socio-technical perspective to explore information reliability and social interactivity as two plausible underlying factors driving the observed changes. Our mechanism exploration reveals that the AIGC ban spurs question volume in topics where AIGC is less reliable and where social interaction is highly expected. In contrast, the ban hampers answer efficiency in communities where LLMs are capable of producing reliable answers and where social interactivity is minimal. Additionally, our results indicate the increased human involvement from knowledge seekers and contributors following the ban. They adapt their behavior by posting questions and answers that are more informationally rich and socially engaging. Overall, our findings offer actionable implications for platform managers, community moderators, and policymakers of online Q&A communities.

[1] Department of Management Science and Technology, Harbin Institute of Technology; Harbin, Heilongjiang, China.

[2] Department of Operations and Information Management, University of Wisconsin–Madison; Madison, WI, USA.

[3] Department of Information Systems and Operations Management, HEC Paris; Jouy-en-Josas, France.

[4] Operations, Information and Decisions Department, The Wharton School, University of Pennsylvania; Philadelphia, PA, USA

* These authors contributed equally to this work.

# The Double-edged Effect of Banning AI-generated Content on Online Question-and-Answer Communities: Evidence from Stack Exchange

## 1. Introduction

Generative artificial intelligence (AI) has rapidly transformed digital knowledge production (OpenAI 2023, Brynjolfsson et al. 2025, Noy and Zhang 2023, Eloundou et al. 2024). Large language models (LLMs), such as ChatGPT and Gemini, generate human-like content in response to natural language prompts and increasingly serve as both alternative sources of information and tools for content creation (Burtch et al. 2024; Quinn and Gutt 2025). Online question-and-answer (Q&A) communities, including Stack Exchange and Quora, facilitate large-scale knowledge exchange and rely heavily on human contributions (Goes et al. 2016; Xu et al. 2020). As LLM use proliferates, platform operators face growing challenges in managing AI-generated content (AIGC), which may alter participation dynamics and content production processes. A common governance response has been to ban AIGC. For example, within four months of ChatGPT's release, eighteen Stack Exchange communities prohibited AI-generated answers due to concerns about reliability, plagiarism, and diminished social authenticity. Similar restrictions were adopted in several Reddit communities.

Platform performance in online knowledge exchange depends on active knowledge seeking, knowledge contribution, and contribution efficiency (Phang et al. 2009; Chen et al. 2025). Because generative AI functions as a substitute and cost-reducing technology (Brynjolfsson et al. 2025; Eloundou et al. 2024), it creates distinct value for different platform participants. Consequently, the effects of AIGC bans are theoretically ambiguous and involve critical trade-offs.

On the one hand, knowledge seekers choose between LLMs and online knowledge platforms to meet their information and social needs (Burtch et al. 2024). Restricting AIGC on platforms may replace low-quality AI-generated answers with authentic, credible human-created responses, thereby enhancing informational credibility and social interactivity and motivating knowledge seeking. On the other hand, prohibiting AIGC on online knowledge platforms removes the cost-reducing advantages of using LLMs to assist content creation. Knowledge contribution may be jointly shaped by the negative effect of increased answer-production costs and the positive effect of expanded opportunities to obtain intrinsic or extrinsic benefits from responding to more posted questions, leading to an ambiguous change in contribution volume. Further, removing the cost-reducing advantages of AIGC may slow resolution speed. Overall, banning AIGC may increase engagement while reducing efficiency, creating trade-offs in platform performance. Because online knowledge platforms vary in topic focus and expectations for social engagement, repositioning platforms toward human involvement by banning AIGC is likely to generate heterogeneous impacts across settings.

Existing research examines how LLM adoption reshapes knowledge exchange, documenting declines in question volume and changes in content quality following LLMs' release (Burtch et al. 2024; Quinn and Gutt 2025; Sanatizadeh et al. 2025; Xue et al. 2023). Yet banning AIGC is not simply the mirror image of LLM adoption. Unlike a pre-LLM environment, a ban does not eliminate LLMs as an external, competing knowledge system; rather, platforms must compete with these AI systems by redefining their own relative strengths. Understanding how platform governance policies reshape platform performance under AI-driven competition is increasingly important as generative AI becomes increasingly embedded in the digital ecosystem.

Three gaps motivate our study. First, prior work focuses primarily on engagement levels—particularly the volume of knowledge seeking and contribution—while overlooking contribution efficiency and potential trade-offs in overall system performance (Burtch et al. 2024; Shan and Qiu 2025). Second, most research centers on programming-oriented communities, providing limited insight into community heterogeneity and how topic focus and social norms shape the interaction between generative AI and online knowledge systems (Quinn and Gutt 2025; Sanatizadeh et al. 2025). Third, some studies explain the interaction between generative AI and online knowledge systems from a demand-supply perspective—for example, how changes in knowledge contribution affect knowledge seeking—without fully accounting for how differences in informational reliability and social expectations jointly determine comparative advantages of external AI systems versus human-centered platforms (Wang and Zheng 2024). Consequently, we address the following research questions:

**RQ1:** *What are the effects of banning AIGC on the knowledge seeking in Q&A communities?*

**RQ2:** *What are the effects of banning AIGC on the knowledge contribution and contribution efficiency in Q&A communities?*

**RQ3:** *What is the plausible underlying mechanism driving the observed effects of the AIGC ban?*

Our empirical context is Stack Exchange, a large network of Q&A communities. Following the release of ChatGPT in late November 2022, 18 Stack Exchange communities officially announced bans on AIGC by April 2023. Moderators enforced these bans using a combination of human review and algorithmic detection tools, removing AI-generated content and, in some cases, suspending violating accounts. We exploit this staggered community-level policy adoption to construct a quasi-experimental setting, comparing communities that banned AIGC with those that did not. After matching treated and control communities on observable characteristics, we aggregate data at the community-week level and estimate difference-in-differences (DiD) models to identify the impact of the ban on knowledge exchange outcomes.

We document a trade-off. Following the ban, the question volume increases by 13% on average, yet the share of questions receiving accepted answers within the standard response window declines by 3.3%. The average number of answers per question remains unchanged. These results indicate improved knowledge seeking but declined contribution efficiency post-ban. Importantly, this trade-off is concentrated in non-STEM communities; we observe no significant effects in STEM communities. These findings are robust to parallel-trend tests, alternative estimators addressing staggered treatment timing, alternative outcome measures, and a range of endogeneity checks.

Mechanism analyses reveal that responses depend on two factors: AIGC reliability and social interactivity. Question volume rises in communities where LLMs are less reliable and where social interaction is highly valued. In contrast, contribution efficiency declines where LLMs perform well and interpersonal engagement is less central. Content analyses further show that both askers and contributors adapt to the ban by producing longer, more socially nuanced, and more effortful content, consistent with heightened expectations of human involvement. These adaptations may explain the observed efficiency decline.

Our study contributes to three streams of literature. First, we extend research on generative AI in online knowledge exchange by providing empirical evidence on the double-edged effects of banning AIGC (Burtch et al. 2024; Shang and Qiu 2025). The results are evident only in non-STEM communities. We show that such governance policies reallocate comparative advantage between external AI systems and human-centered knowledge communities, generating divergent effects on knowledge seeking and contribution efficiency. Second, we contribute to the literature on machine substitution and augmentation (Autor and Dorn 2013; Acemoglu and Restrepo 2018; Dixon et al. 2021) by demonstrating that AI substitution in digital human-centered systems depends not only on task routineness but also on informational reliability and social expectations. More broadly, our study advances literature on platform competition and differentiation (Cennamo and Santalo 2013; Rietveld and Schilling 2021), providing a socio-technical account of how digital platforms adapt to rising generative AI adoption and competition. As generative AI becomes widely accessible, platforms that rely on human contributions must reposition along socio-technical dimensions where human involvement provides distinct value, while remaining attentive to the system performance trade-offs such repositioning may entail.

## 2. Related Work

### 2.1 Knowledge Exchange and Platform Performance

Knowledge exchange is central to organizations and digital platforms (Ko et al. 2005; Bock et al. 2005). Prior research distinguishes between knowledge seeking and knowledge contribution, showing that participation is shaped by information quality, expected learning benefits, incentives, reputation concerns, and social embeddedness (Goes et al. 2016; Xu et al. 2020; Chen et al. 2018; Kuo and Lee 2009; Wasko and Faraj 2005; Wang et al. 2022).

While this literature provides important insights into the drivers of participation, it focuses primarily on engagement levels, such as the volume of knowledge seeking and contribution, rather than overall system performance. In particular, contribution efficiency, which captures the timeliness and effectiveness of knowledge resolution, remains largely understudied (Chen et al. 2025). As platforms increasingly compete with external AI systems such as LLMs, evaluating system performance requires understanding not only participation but also trade-offs between engagement and efficiency.

## 2.2 Generative AI as a Substitute and Cost-Reducing Technology

With the rise of generative AI, a growing body of research has examined how LLMs affect online knowledge exchange. Appendix E summarizes this literature. This stream implicitly conceptualizes generative AI as either a substitute for, or a cost-reducing technology in, online knowledge exchange. On the knowledge-seeking side, LLMs function as substitute information sources that divert user attention away from traditional human-centered platforms. For example, Burtch et al. (2024) document declines in question volume on Stack Overflow following ChatGPT's release. On the knowledge-contribution side, LLMs act as cost-reducing tools that enhance contributor productivity and alter content characteristics, improving readability, novelty, or complexity of contributed answers on knowledge exchange platforms (Quinn and Gutt 2025; Shan and Qiu 2025; Sanatizadeh et al. 2025), though with mixed effects on community evaluation metrics such as votes that questions and answers receive (del Rio-Chanona et al. 2024).

Several recent studies also reveal substitutive and cost-reducing roles of generative AI by examining platform governance policies on AIGC use in online knowledge platforms. For example, Wang and Zheng (2024) find overall declines in both question and answer volume in Stack Overflow-dominated online communities following the AIGC ban. Their results indirectly suggest that LLMs reduce the cost of content creation. They further adopt a demand-supply perspective, explaining that the post-ban reduction in answer volume (supply) partly drives the decline in question volume (demand). Beyond engagement levels, Borwankar et al. (2023) document post-ban changes in content tone and complexity.

These recent studies primarily focus on the volume of knowledge seeking and contribution, potentially overlooking important trade-offs in evaluating the consequences of AIGC adoption and platforms' response strategies. Moreover, they rely heavily on programming-oriented communities, leaving broader community heterogeneity underexplored. Further, much of this work remains grounded in participation-based and demand–supply explanations. While informative, these perspectives do not fully capture how generative AI interacts with human-centered knowledge exchange.

### 2.3 Competition Between Generative AI and Online Knowledge Exchange

Classic machine substitution theory emphasizes task routineness and codifiability as key determinants of automation (Autor and Dorn 2013; Acemoglu and Restrepo 2018). In this view, machines substitute for humans where tasks are structured and objective. However, an extended lens is needed to understand the interaction between generative AI and online knowledge exchange. Beyond traditional task execution, online knowledge exchanges operate as a socio-technical system in which value creation depends jointly on informational reliability and social interaction (Phang et al. 2009). Moreover, unlike traditional automation technologies, generative AI is versatile across a wide range of tasks, yet its reliability varies across domains (Eloundou et al. 2024). It also exhibits the capacity to mimic human interactions, shaping perceived social engagement (Arora et al. 2025). Therefore, the competition between generative AI and online knowledge exchange unfolds along both informational and social dimensions.

By adopting a socio-technical perspective, we shift the analytical focus from task routineness to differences in reliability and social engagement when examining the consequences of AIGC bans on online knowledge platforms. This perspective implies that generative AI systems outperform human-centered platforms where AIGC is reliable and social interaction is less central, whereas human contributors retain advantages in domains requiring contextual nuance and interpersonal engagement. Consequently, banning AIGC may produce divergent effects across settings. Our study develops and empirically tests this socio-technical comparative advantage framework in the context of AIGC bans.

## 3. Theoretical Considerations

### 3.1 Online Q&A Communities as a Socio-technical System

Online Q&A communities are socio-technical systems in which value creation depends jointly on technical and social capabilities (Phang et al. 2009). Technical capabilities refer to the extent to which systems enable individuals to fulfill their knowledge-seeking and contribution needs easily and reliably. By contrast, social capabilities refer to the extent to which systems support interpersonal interactions, maintain community norms, and facilitate the achievement of individuals' social goals or the community's shared purposes.

The socio-technical perspective underscores that the activities of knowledge seekers and contributors are motivated by the satisfaction of both information-related and social needs. Knowledge seeking and contribution are shaped not only by the accuracy and accessibility of information, but also by social interactivity (Liu et al. 2025; Liao et al. 2024). This perspective has been widely applied in prior literature and is supported by extensive empirical evidence. For example, Liu et al. (2025) examine cross-department knowledge contribution in corporate online communities and reveal knowledge-focused (technical) and social-focused (social) needs as two underlying motives driving knowledge contribution. Similarly, Liao et al. (2024) and Wu (2013) conceptualize two types of community capabilities: knowledge access and repository functions (technical), and the facilitation of user-to-user social interactivity (social), empirically showing that both positively affect knowledge contribution and productivity.

The socio-technical perspective also aligns with two central dimensions and concerns about generative AI and AIGC amid their growing adoption across various domains (Eloundou et al. 2024). On the technical side, numerous studies report that AIGC exhibits varying levels of information reliability across topics and tasks (Mahaut et al. 2024, Wang et al. 2024). On the social side, although generative AI can mimic individual behaviors and simulate interpersonal interactions in tasks traditionally reliant on humans (Horton 2023; Arora et al. 2025), it still provides only limited social experience (Meng et al. 2025).

Drawing upon the socio-technical framework and extant literature, generative AI differs from online knowledge exchange communities in its relative strengths across informational and social dimensions. Banning AIGC from online Q&A communities does not eliminate generative AI as a substitute knowledge source or cost-reducing tool; rather, it reshapes online communities' comparative advantages along socio-technical dimensions. We therefore adopt a comparative advantage perspective: the AIGC ban potentially increases engagement in domains where humans hold a comparative advantage while reducing efficiency in areas where generative AI previously enhanced productivity.

### 3.2 AIGC Ban and Knowledge Seeking

Knowledge seekers choose between online Q&A communities and external AI systems such as LLMs to satisfy informational and social needs (Phang et al. 2009; Liu et al. 2025). The emergence of LLMs provides an alternative source for obtaining immediate responses, particularly for common questions that these models can address reliably (Burtch et al. 2024; del Rio-Chanona et al. 2024). The availability of LLMs also influences the types and quality of answers that knowledge seekers receive from communities, as contributors may post AI-generated answers alongside human-generated responses, and answer quality may also be affected by the presence of such content (Quinn and Gutt 2025; Sanatizadeh et al. 2025). As a result, knowledge seeking reflects a comparative choice between LLMs and online knowledge platforms.

The AIGC ban promotes human involvement in the answer contribution while keeping LLMs as an alternative resource for knowledge seekers. From a comparative advantage perspective, banning AIGC alters the platform's relative position along technical and social dimensions. First, although LLMs can provide correct responses for a wide range of tasks (Noy and Zhang 2023), they may also hallucinate answers. Restricting AIGC enhances the platform's informational credibility, particularly in communities where the AI reliability is perceived to be low, such as those featuring context-dependent, nuanced, or experience-based questions. By emphasizing human accountability and effort, the ban strengthens the platform's technical advantage in domains where AI reliability is limited. In such contexts, knowledge seeking is more likely to increase following the ban.

Second, although LLMs can generate human-like content and conversational responses, AIGC generally lacks authentic human experience and efforts, providing limited meaningful social engagement (Kirk and Givi 2025). Banning AIGC reinforces the platform's social distinctiveness, particularly in communities where social interaction is central to value creation. By prioritizing human-generated answers and enforcing human effort in answer creation, the ban enhances opportunities for meaningful interpersonal engagement. Accordingly, knowledge seeking is more likely to rise, especially in communities where social interactivity is highly valued.

### 3.3 AIGC Ban and Knowledge Contribution

Knowledge contributors participate when perceived benefits exceed production costs (Phang et al. 2009; Wu et al. 2023). LLMs and AIGC reduce the cognitive and execution costs of knowledge contribution by assisting with information retrieval, synthesis, and drafting (Sanatizadeh et al. 2025; Shan and Qiu 2025).

Banning AIGC removes this cost advantage. The ban requires contributors to engage in a more effortful process of information search and content creation, which may discourage some from posting answers. Without access to AIGC trained on large-scale datasets, contributors also lose access to a rich knowledge repository, potentially undermining their efficacy and information-driven motivations to contribute (Liu et al. 2025).

However, contribution responses may vary across contexts. If banning AIGC increases question volume in domains where human expertise retains a comparative advantage, contributors may perceive greater opportunity for intrinsic or extrinsic gains, motivating them to post more answers. Moreover, restricting AIGC may reduce reward exploitation using low-effort AI-generated answers, thereby strengthening perceived fairness on the platform and incentivizing answer contributions, particularly from experienced contributors (Li and Kim 2025). Thus, the net effect of banning AIGC on contribution volume is theoretically ambiguous and likely contingent on the platform's comparative advantage relative to LLMs.

### 3.4 AIGC Ban and Contribution Efficiency

Contribution efficiency reflects the speed and effectiveness with which questions are resolved. LLMs substantially reduce the time required to retrieve information, synthesize content, and draft responses (Noy and Zhang 2023). From a socio-technical comparative advantage perspective, the AIGC ban may have heterogeneous influences on contribution efficiency. In domains where AI reliability is high and social interaction is less central, contributors can leverage AI tools to generate acceptable answers rapidly, conferring a clear technical efficiency advantage. Banning AIGC removes this advantage, forcing contributors to rely solely on human effort and increasing the time required to produce satisfactory responses. Contribution efficiency is therefore likely to decline. By contrast, in contexts with low AIGC reliability and high expectations for social engagement, human effort already dominates the creation of high-quality answers. In these settings, promoting human involvement has little impact on existing content creation practices, and efficiency is likely to remain stable post-ban.

In addition, repositioning the platform toward human-centered knowledge exchange may reshape user expectations and promote norm compliance. By emphasizing authenticity and human effort in online knowledge platforms, the ban may raise knowledge seekers' expectations for human-provided answers, encouraging them to post more complex or socially nuanced questions. Contributors may correspondingly adapt by providing more detailed and effortful answers to meet these heightened norms. This norm-driven adjustment can further extend resolution time, reinforcing declines in contribution efficiency.

In summary, the socio-technical comparative advantage framework yields three core predictions. First, banning AIGC may increase knowledge seeking, particularly in domains where AI reliability is limited or social interaction is highly valued. Second, its effect on contribution volume is theoretically ambiguous, depending on the removal of cost advantage and shifts in question volume. Third, contribution efficiency is likely to decline where AI previously enhanced productivity in producing high-quality answers, especially in contexts with high information reliability and limited social interaction. These predictions reflect a structural trade-off rather than a purely negative outcome and highlight systematic heterogeneity in platform responses to AI governance policies.

## 4. Research Setting and Data

### 4.1 Research Setting

We use Stack Exchange, a prominent network of Q&A communities, as our research context to empirically examine the consequences of the AIGC ban on knowledge seeking, knowledge contribution, and contribution efficiency in online Q&A communities. As of April 2023, Stack Exchange comprises nearly 180 communities, spanning a wide range of topics, such as technology, science, culture, and the arts. Users participate by asking questions and contributing answers. These communities rely heavily on volunteer moderators, who play a crucial role in overseeing content creation and community governance. These moderators actively engage with community members and frequently update community guidelines to maintain content quality and foster a productive community culture.

Generative AI-created content and users' discussions mentioning generative AI were observed across Stack Exchange communities at least one month before the release of ChatGPT, as detailed in Appendix A1 and documented by Shan and Qiu (2025). Following the launch of ChatGPT on November 30, 2022, there was a surge in AIGC activity along with discussions. Although users were not required to explicitly label AIGC, these discussions indicate widespread awareness of AIGC within the communities, as well as user concerns about the reliability of AIGC and the potential loss of social value derived from interacting with real human users due to its use. For example, the English Language & Use community raised issues about answer reliability and fairness associated with AIGC.[5] In response to similar concerns, moderators across various communities announced official AIGC bans at different points in time. These communities adopted algorithmic detectors along with moderator judgment to identify and remove AIGC, and users who violated these policies risked suspension or permanent bans. While detection mechanisms cannot reach full accuracy, proactive enforcement by moderators combined with the potential consequences of rule violations signals a clear shift toward human-created content and likely discourages users from submitting AIGC. Pre-existing awareness and concerns about AIGC before the ban suggest that users may adjust their behavior in response to signaled changes, likely leading to changes in the activities of knowledge seekers and contributors, and consequently affecting knowledge exchange outcomes.

Eighteen communities had implemented an AIGC ban as of April 2023. Appendix A2 provides a complete list of these communities along with the dates of their announcements. We use communities that implemented the AIGC ban as the treatment group in our study, with the ban announcement date serving as the treatment implementation date. Communities that had not banned AIGC by April 2023 constitute the control group. This setup forms a natural experiment, allowing us to examine the effects of the AIGC ban.

**4.2 Preliminary Evidence for the Effectiveness of the AIGC Ban**

Our observational window is from November 30, 2022, to April 8, 2023. Before conducting the main analyses, we first collect descriptive evidence on the change in the volume of plausibly AI-generated content on Stack Exchange and assess whether the AIGC ban effectively reduces such content. Since Stack Exchange does not explicitly label content (e.g., a posted answer) as AI- or human-generated, we employ a widely applied detection algorithm to assess the likelihood that a given answer was generated by AI (Guo et al. 2023; Liu et al. 2019; Xu and Sheng 2024). These tools are designed to identify content produced by ChatGPT and its precursors such as GPT-2 and GPT-3. Given that AIGC is mostly used in answer creation rather than questions (Shan and Qiu 2025), we apply the AIGC detection algorithm to all answers posted during the observational period. Each answer is assigned a likelihood score of being AI-generated, and we label an answer as AIGC if the likelihood exceeds 90%.

[5] See https://english.meta.stackexchange.com/questions/15500/announcement-ai-generated-answers-are-officially-banned-here (last accessed on February 28, 2026).

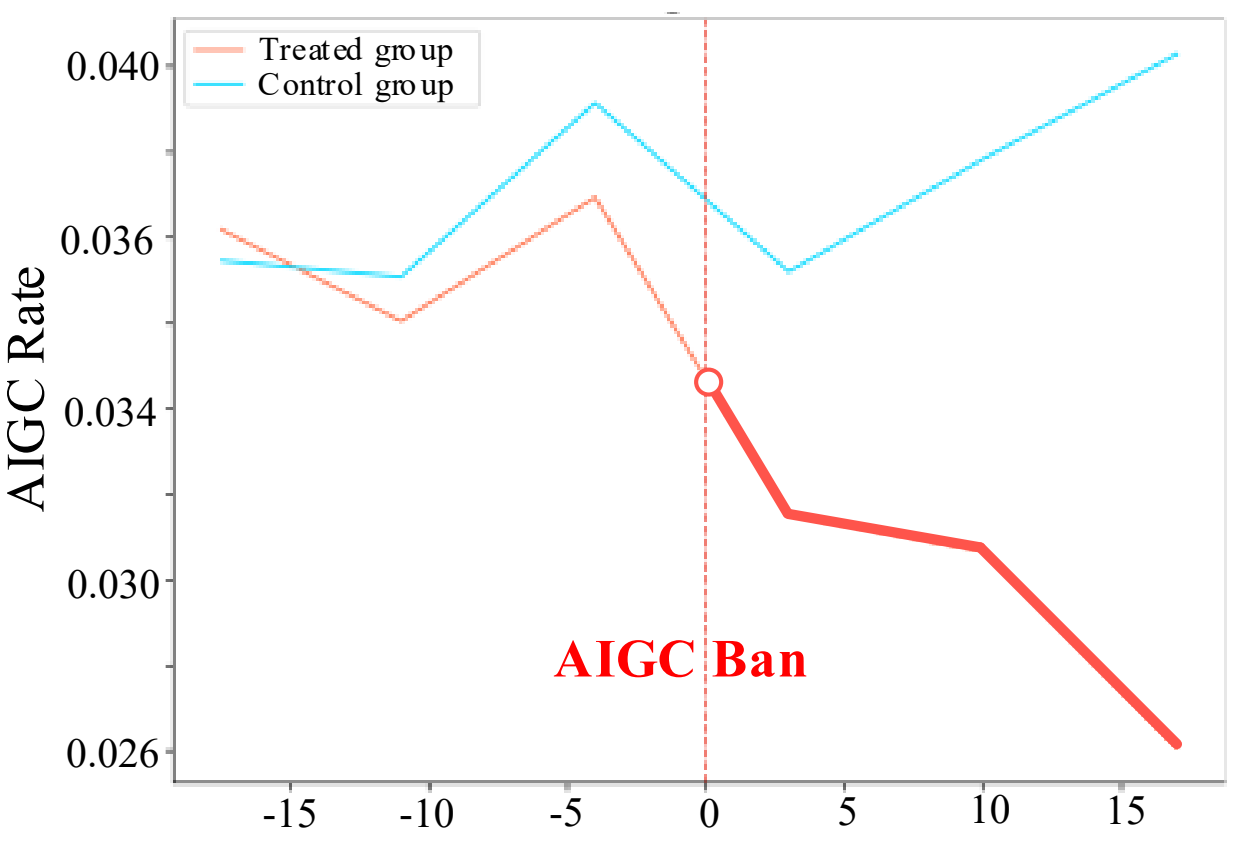

**Figure 1. AIGC Rate in Treated and Control Communities**

Note: Treated communities with different banning dates are normalized to the same timeline based on the number of days following the AIGC ban.

Among the 230 thousand answers in our sample, 3.6% are detected as AIGC. Comparing AIGC rates between treated and control communities reveals distinct post-ban patterns. Figure 1 shows that the AIGC rates in both groups were comparable before the ban (*t*-statistics = 0.247, p>0.1), but diverged afterwards. Specifically, the AIGC rate in the control group slightly increases, whereas the rate in the treatment group sharply decreases. A *t*-test confirms that the post-ban differences between these groups are statistically significant (*t*-statistic = 3.451, p<0.01). These findings suggest that although AIGC continued to appear even after the ban—reflecting the practical challenges moderators face in fully detecting and eliminating such content—the overall prevalence of AIGC in treated communities is significantly lower than in control communities. This provides empirical support for the effectiveness of the AIGC ban in curbing the AIGC.

### 4.3 Identification Strategy

Given the variation in community-level policies toward AIGC, we employ a difference-in-differences (DID) design combined with community matching to estimate the causal impact of the AIGC ban on knowledge seeking, knowledge contribution, and contribution efficiency in online Q&A communities. We collect user activity data from November 30, 2022 to April 8, 2023 using the Stack Exchange Data Explorer. The dataset covers all user activities in the 18 communities that implemented an AIGC ban (treatment group) and in the remaining communities that had not imposed such a ban as of April 2023 (control group).

To ensure comparability between treated and control communities, we match communities on a series of observable characteristics that may influence a community's decision to ban AIGC. These characteristics are constructed using data from November 1 to December 5, 2022—a period before the first AIGC ban announcement on Stack Exchange. Specifically, we match communities on the following aspects: (1) number of months a community had been launched as of December 2022 (CommunityAge); (2) average number of questions per day (Questions); (3) average number of answers per question (AnswersPerQ); (4) proportion of questions receiving at least one answer (AnswerRate); (5) the number of deleted answers (AnswerDeletion); (6) average number of views per post (AvgViews); (7) average answer score (AvgScore); and (8) whether the community belongs to a STEM or a non-STEM domain (STEM). Recent econometric advances highlight both the benefits and costs of matching on pre-treatment outcomes in DiD analyses and the importance of evaluating the appropriateness of including such variables (Chabé-Ferret 2017; Ham and Mitatrix 2024). As detailed in Appendix B1, we perform formal tests on matching covariates constructed from pre-treatment outcomes, and the results support their inclusion in our matching procedure.

Table B2 in Appendix B.2 presents the summary statistics of all matching covariates. We apply kernel propensity score matching, resulting in a matched sample comprising 15 treated communities and 85 control communities. Table B3 provides the results of the balance test, demonstrating that the matched communities are statistically comparable across the specified community-level dimensions.

In addition to general community characteristics, AIGC-related factors, such as the prevalence of AIGC and the extent of community concerns surrounding its use, may drive communities' decision to ban AIGC. Although we did not explicitly match on these AIGC-related characteristics, the selected matching covariates are likely correlated with them, allowing the matching procedure to identify treated and control communities that are comparable along these dimensions. We, therefore, conduct additional balance tests on several AIGC-related characteristics in the matched sample, such as the AIGC rate, the number of AIGC-related discussions, and the average reputation scores of users posting AIGC. Summary statistics and balance test results for these additional characteristics are presented in Appendix B.3. Although some differences between treated and control communities are observed before the matching, these differences are eliminated in the matched samples. These additional checks further strengthen the causal interpretation of results reported in the subsequent analyses.

### 4.4 Outcome Variables

We organize the data at the community-week level and construct variables to proxy for our outcomes of interest. Table 1 reports detailed descriptions and summary statistics for these variables. Specifically, we use the number of questions posted in a community each week (*NumQuestion*) to measure knowledge seeking. We measure knowledge contribution using the average number of answers per question in a given week (*NumAnswerPerQ*), which accounts for variation in question volume and helps avoid the 'bad control' issue in the regression estimation.[6] In the empirical analyses, we apply a log transformation after adding one to each measure to address the skewed distribution of these two variables and accommodate zero values.

**Table 1. The Summary Statistics of Outcome Variables**

| Variables | Description | Mean | S.D. | Min | Max |
|---|---|---|---|---|---|
| $Log(NumQuestion_{it})$ | The number of questions in community *i* in week *t*. | 2.784 | 1.423 | 0 | 6.265 |
| $Log(NumAnswerPerQ_{it})$ | The average number of answers per question in community *i* in week *t*. | 0.679 | 0.324 | 0 | 1.887 |
| $WithAcceptedAsw_{it}$ | The proportion of questions that have an accepted answer within eight hours following the question creation in community *i* in week *t*. | 0.163 | 0.158 | 0 | 1 |

[6] A 'bad control' issue arises when empirical models include control variables that are affected by the treatment, which can bias the estimated treatment effects (Angrist and Pischke 2009).

To capture contribution efficiency, we utilize the proportion of questions that receive an accepted answer within a specified time window after the question is posed (*WithAcceptedAsw*) (Chen et al. 2025). On Stack Exchange, each asker, upon receiving answers, can select only one answer as "accepted," signaling that the question has been satisfactorily resolved. On average, accepted answers are posted within eight hours, and over 70% of questions with accepted answers fall within this time frame. Therefore, we consider eight hours as the expected time to receive a satisfactory answer on the platform, and the measure constructed based on this time window serves as a valid proxy for contribution efficiency.

## 5. Empirical Model and Results

### 5.1 Empirical Model

With the matched treatment and control communities, we employ a difference-in-differences (DID) approach to estimate the impact of the AIGC ban on knowledge seeking, knowledge contribution, and contribution efficiency. Our empirical model is specified in Equation (1):

$$DV_{it} = \beta_0 + \beta_1 AIGC_Ban_{it} + x_i + \theta_t + \varepsilon_{it} \quad (1)$$

where the dependent variable $DV_{it}$ represents three outcomes of interest, measured by Log($NumQuestion_{it}$), Log($NumAnswerPerQ_{it}$), and *WithAcceptedAswit*, respectively. $AIGC_Ban_{it}$ is a dummy variable indicating whether community $i$ has an active AIGC ban in week $t$. The coefficient $\beta_1$ captures the effect of the AIGC ban and is the main estimate of interest. We include community fixed effects ($x_i$) and week fixed effects ($\theta_t$) to control for unobserved time-invariant community characteristics and common temporal factors. $\varepsilon_{it}$ is the idiosyncratic error term. We apply the robust standard errors clustered at the community level in the estimation.

### 5.2 Main Results

We report the main results in Table 2. On average, the number of questions posted in a community increases by 13% following the AIGC ban, indicating that emphasizing human involvement in online Q&A communities motivates users to ask more questions. As for knowledge contribution, we find no significant change in the average number of answers per question[7]. However, the estimate reported in Column (3) of Table 2 shows a decline in the proportion of questions receiving accepted answers within the expected timeframe, suggesting reduced contribution efficiency in a community after the ban.

Our results highlight the double-edged effects of the AIGC ban. While emphasizing a human-centered community and promoting human-generated content boosts question volume, the ban simultaneously reduces answer efficiency. In the short term, this policy appears to encourage greater engagement from knowledge seekers by stimulating question-asking behavior. However, the accumulation of unanswered and delayed questions may become a bottleneck over time, potentially hindering user engagement and satisfaction.

**Table 2. Main Results**

| | (1) | (2) | (3) |
|---|---|---|---|
| | Log(NumQuestion) | Log(NumAnswerPerQ) | WithAcceptedAsw |
| $AIGC_Ban_{it}$ | 0.130*** | -0.006 | -0.033** |
| | (0.047) | (0.027) | (0.015) |
| Constant | 2.769*** | 0.680*** | 0.167*** |
| | (0.005) | (0.003) | (0.002) |
| No. Observations | 1,900 | 1,831 | 1,831 |
| Community Fixed Effects | Y | Y | Y |
| Week Fixed Effects | Y | Y | Y |

Note: Robust standard errors are clustered at the community level and reported in parentheses. * $p$<0.1; ** $p$<0.05; *** $p$<0.01.

### 5.3 Heterogeneity Results

We leverage the rich heterogeneity across communities in our dataset to examine how the effects of the AIGC ban vary by community type. Specifically, we distinguish between STEM and non-STEM communities, as the desirable content and expected interactions differ fundamentally across these domains. In STEM communities, questions typically pertain to objective topics with clear and verifiable answers grounded in established scientific principles and evidence. Such questions often involve varying levels of complexity and may require specialized knowledge for accurate resolution. In contrast, non-STEM communities more frequently feature subjective, context-dependent discussions that emphasize diverse opinions and authentic human experiences. Given these differences, knowledge seekers may perceive varying values from AIGC versus human-created content, and knowledge contributors may benefit differently from the availability of AIGC across topic domains. Accordingly, we conduct subgroup analyses to empirically examine how the impact of the AIGC ban varies between STEM and non-STEM communities.

[7] In addition to examining changes in the number of answers per question following the AIGC ban, we analyze changes in the total answer volume within a community post-ban. The results show an increase in total answer volume in non-STEM communities, likely driven by a rise in question volume. Detailed analyses and results are provided in Section 5.4.3 and Appendix D.1.

Operationally, we classify communities as STEM or non-STEM based on Stack Exchange's categorization. Specifically, communities focused on technology and science are designated as STEM, while those centered on life and art, culture, and professionals are categorized as non-STEM. We re-estimate the effect of the AIGC ban separately for each community type and report the results in Table 3. Interestingly, the double-edged effect of the AIGC ban emerges only in non-STEM communities. Specifically, the number of questions posted in non-STEM communities increases following the ban, whereas those in STEM communities remain unchanged. A possible explanation is that knowledge seekers in non-STEM communities place greater value on authentic, human-created answers than their counterparts in STEM communities. By emphasizing human involvement, the AIGC ban enhances knowledge seekers' perceived value of online Q&A communities for non-STEM topics, stimulating question-asking behavior.

Additionally, non-STEM communities experienced a decline in answer efficiency after the ban, while answer efficiency in STEM communities remains unchanged. This discrepancy may stem from differences in the ease of generating plausible answers across domains. When AIGC is allowed, answer contributors can rely on LLMs to quickly generate reasonable responses to non-STEM questions, which often involve subjective interpretation and diverse opinions. By contrast, producing satisfactory answers to STEM questions typically requires factual evidence and complex logical reasoning, making reliance on LLMs (at least. during our research sampling period) less effective even when AIGC is allowed.[8] Consequently, removing AIGC disproportionately reduces answer efficiency in non-STEM communities,

[8] Of course, over time, as reasoning models improve, the discrepancy of LLM performance between STEM and non-STEM may shrink. At least during our sample, producing cogent logical reasoning was difficult for LLMs.

where contributors must invest greater human effort and time to produce reliable answers post-ban. In STEM communities, however, answer contributors primarily rely on human effort to produce or check answers regardless of AIGC availability, resulting in little change in answer efficiency following the ban.

**Table 3. STEM vs. non-STEM Communities**

| | (1) Log(NumQuestion) | (2) Log(NumAnswerPerQ) | (3) WithAcceptedAsw |
|---|---|---|---|
| | Panel A: STEM Communities | | |
| $AIGC_Ban_{it}$ | 0.06 | -0.04 | -0.007 |
| | (0.078) | (0.043) | (0.015) |
| Constant | $3.156^{***}$ | $0.569^{***}$ | $0.144^{***}$ |
| | (0.01) | (0.006) | (0.002) |
| No. Observations | 1,064 | 1,042 | 1,042 |
| | Panel B: non-STEM Communities | | |
| $AIGC_Ban_{it}$ | $0.224^{***}$ | 0.026 | $-0.067^{**}$ |
| | (0.046) | (0.03) | (0.025) |
| Constant | $2.281^{***}$ | $0.829^{***}$ | $0.195^{***}$ |
| | (0.004) | (0.003) | (0.002) |
| No. Observations | 836 | 789 | 789 |
| *p-value* of the permutation test (500 times bootstrap) | $0.026^{**}$ | $0.092^{*}$ | $0.03^{**}$ |
| Community Fixed Effects | Y | Y | Y |
| Week Fixed Effects | Y | Y | Y |

Note: Robust standard errors are clustered at the community level and reported in parentheses. * $p<0.1$; ** $p<0.05$; *** $p<0.01$.

Given that STEM and non-STEM communities differ in their informational focus and expectation of social interactions, the observed heterogeneity across these community types likely reflects the combined influences of the AIGC ban along both informational and social dimensions. Guided by the theorization in Section 3, we conduct further analyses in Section 6 to disentangle the underlying mechanisms.

### 5.4. Robustness Checks

We validate our main findings using an extensive set of robustness checks, as detailed in the following subsections and Appendix C. Overall, the results remain robust, reinforcing the conclusion that the AIGC ban imposes a significant trade-off for Q&A communities.

**Table 4. Summary of Robustness Check**

| Concerns | Adopted Methods to Address Concerns | Results |
|---|---|---|
| Validity of DiD estimates | Parallel trend test | Figure 2 |
| Community characteristics as confounders | Test on matching covariates constructed based on pre-treatment outcomes | Appendix B.1 |
| | Test on general community characteristics | Appendix B.2 |
| | Test on AIGC-related community characteristics | Appendix B.3 |
| Measurement validity | Alternative measures of contribution efficiency | Appendix C.1 |
| Unobserved, time-varying confounders | Counterfactual estimators | Appendix C.2 |
| | Placebo test | Appendix C.4 |
| | Control for exposure to other AIGC bans | Appendix C.6 |
| Time-invariant unobservables and latent trends | Look-ahead matching | Appendix C.5 |
| Heterogeneous treatment effects and biased TWFE estimates under the staggered treatment adoption | Callaway and Sant'Anna (2021) estimator | Table 5 |
| | Counterfactual estimators | Appendix C.2 |
| | Two-staged DiD estimator | Appendix C.2 |
| Short pre-treatment period | Pre-existing AIGC awareness and its driving effects | Appendices A.1 & D.5 |
| | Alternative estimators | Section 5.4.2 Appendix C.2 |
| | Exclusion of communities that banned AIGC early | Appendix C.3 |

### 5.4.1 Parallel Trends Assumption

Given that our research employs the DID as the main identification framework, we validate our results by examining the parallel trends assumption required for the DID estimations. Specifically, we implement the relative time model by replacing the dummy variables $AIGC_Ban_{it}$ in Equation (1) with $RelativeWeek_{it}$, as specified in Equation (2). The variable $RelativeWeek_{it}$ indicates the temporal distance of each observation from the implementation of the AIGC ban in weeks.

$$DV_{it} = \beta_0 + \beta_1 RelativeWeek_{it} + x_i + \theta_t + \varepsilon_{it} \quad (2)$$

We use the week preceding the implementation of the AIGC ban as the reference week. Figure 2 visualizes the estimates from the relative time model. The differences between the treatment and control groups— particularly in the weeks leading up to the policy implementation—are insignificant at the 5% significance level, providing supportive evidence for the parallel trends assumption. Although several weeks prior to week -8 show significant differences between the matched treated and control communities in terms of question volume, this is less concerning because these cases occur much earlier than the typical pre-treatment period (approximately 5 weeks) in the dataset and do not exhibit a consistent pattern of violation of the parallel trend assumption. We observe stronger evidence of parallel trends when visualizing the results for non-STEM communities, strengthening confidence in our estimates.

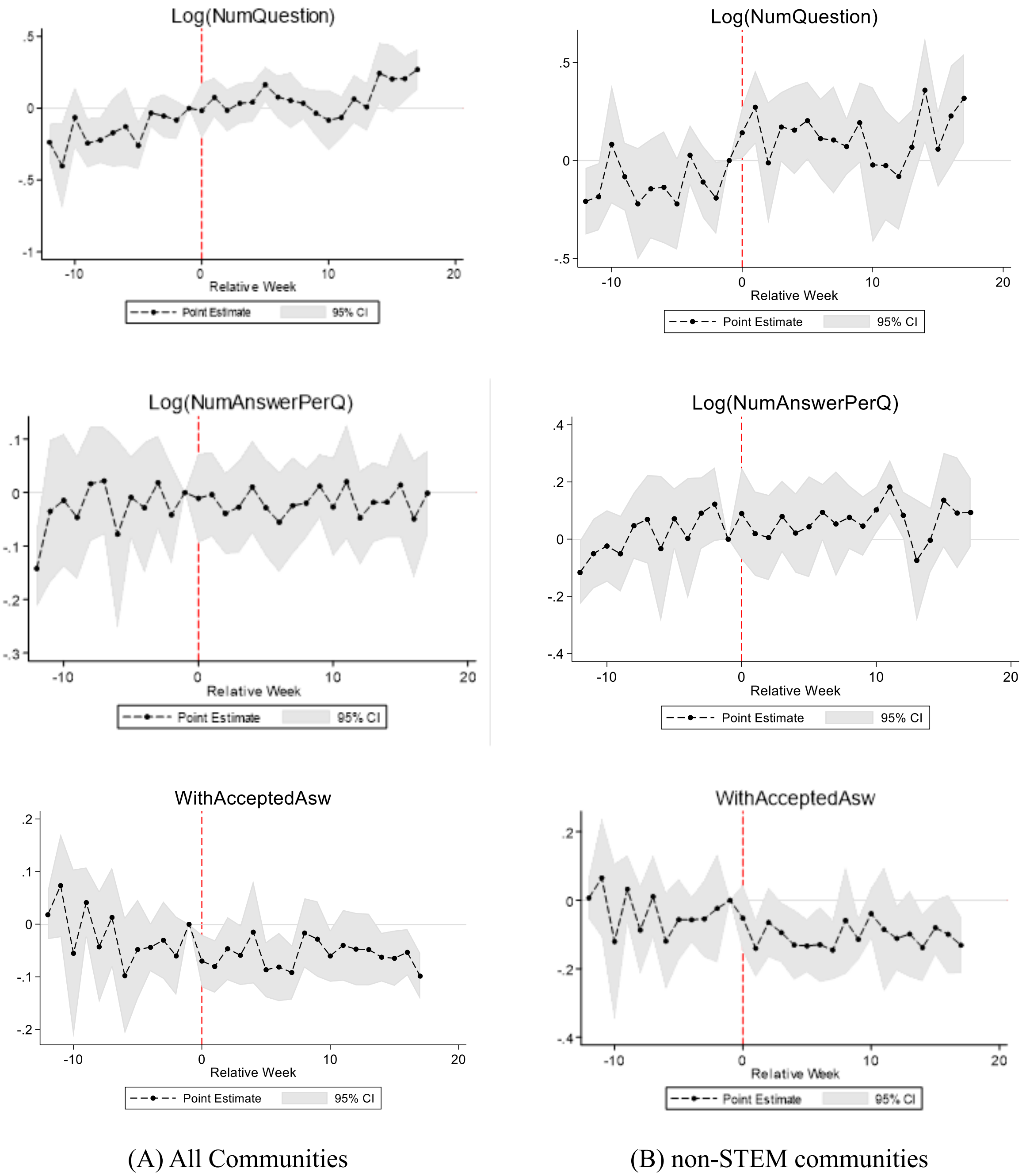

**Figure 2. Results of the Relative Time Model**

### 5.4.2 Alternative Estimators

In our context, the AIGC ban was implemented at different times across Stack Exchange communities. Recent advances in econometric methods suggest that staggered treatment adoption may result in heterogeneous treatment effects across units (Goodman-Bacon 2021) and that applying a two-way fixed effects approach in such contexts may lead to biased estimates. Accordingly, we validate our results using several recent methods designed to produce robust estimates in empirical settings with variation in treatment timing (i.e., staggered entry) and to appropriately account for heterogeneous treatment effects.

Specifically, we employ the estimators proposed by Callaway and Sant'Anna (2021), using communities that have not yet implemented the AIGC ban as the control group to estimate the ban's impact on three outcomes of interest. Table 5 reports the results. Across all communities, we find no statistically significant effect of the AIGC ban. Notably, while restricting the analyses to non-STEM communities, the results in Panel C indicate that the AIGC ban positively affects knowledge seeking but negatively affects contribution efficiency. These findings are consistent with our earlier observations on community-level heterogeneity and confirm the double-edged effects of the AIGC ban in non-STEM communities.

**Table 5. Results of Alternative Estimators (Callaway and Sant'Anna 2021)**

| | (1)<br>Log(NumQuestion) | (2)<br>Log(NumAnswerPerQ) | (3)<br>$WithAcceptedAsw_{it}$ |
|---|---|---|---|
| | Panel A: All Communities | | |
| $AIGC_Ban_{it}$ | 0.070<br>(0.069) | 0.046<br>(0.046) | -0.023<br>(0.032) |
| No. Observations | 1,862 | 1,793 | 1,793 |
| | Panel B: STEM Communities | | |
| $AIGC_Ban_{it}$ | -0.001<br>(0.099) | 0.059<br>(0.066) | 0.038<br>(0.028) |
| No. Observations | 1,026 | 1,004 | 1,004 |
| | Panel C: Non-STEM Communities | | |
| $AIGC_Ban_{it}$ | 0.168**<br>(0.081) | 0.023<br>(0.06) | -0.114**<br>(0.05) |
| No. Observations | 836 | 789 | 789 |
| Community Fixed Effects | Y | Y | Y |
| Week Fixed Effects | Y | Y | Y |

Note: Robust standard errors are clustered at the community level and reported in parentheses; * $p<0.1$; ** $p<0.05$; *** $p<0.01$; when applying the Callaway and Sant'Anna estimator, some observations that do not satisfy the estimator's structural requirements are automatically excluded.

We also applied two alternative approaches that are well-suited to settings with staggered treatment adoption to further validate our results. First, we apply counterfactual estimators (Liu et al. 2024; Athey et al. 2021), which produce robust estimates by directly imputing counterfactual outcomes for treated communities and are effective in environments with heterogeneous treatment effects and unobserved time-varying confounders. Second, we employ the two-staged difference-in-differences method proposed by Gardner (2022), which produces robust estimates through a two-stage estimation framework based on generalized method of moments estimators. Overall, our results remain robust. We present the analysis details and results in Appendix C.2.

#### 5.4.3 Additional Robustness Tests

We conduct additional robustness checks to validate our findings. We start with alternative measurements. In the main analyses, we measure contribution efficiency as the proportion of questions that received accepted answers within eight hours of question creation. We validate the results by (i) using alternative time thresholds for the efficiency measurement, and (ii) adding additional control variables. Appendix C.1 reports the details of these tests, and our results remain consistent. Additionally, our main analysis measures knowledge contribution using the average number of answers per question in a community. While this measure accounts for changes in question volume in a community and avoids the 'bad control' issue in the estimation, it does not capture changes in the overall knowledge contribution. We thus supplement the main analyses by estimating post-ban changes in total answer volume. We find that total answer volume in non-STEM communities increases following the ban; however, this increase becomes insignificant once we control for the rise in question volume. This finding is consistent with the main results showing no change in answer contribution per question. Appendix D.1 reports the detailed analyses and results reported.

We also empirically test and address several concerns related to the swift implementation of the AIGC ban in some treated communities. First, the close temporal proximity between the release of ChatGPT and the ban raises concerns about limited and less-established AIGC activity before the treatment. However, as illustrated in Appendix A.1, AIGC (using precursors of ChatGPT such as GPT-3) and related concerns were already frequently present in the month preceding ChatGPT's launch, suggesting a longer period of AIGC prevalence and well-established awareness before the ban. This awareness plausibly drives changes in the activities of knowledge seekers and contributors, as evidenced by results in Appendix D.5, which show the observed changes are significant only in communities with high AIGC awareness. Second, we replicate the analyses by restricting the treatment group to communities with at least one month of pre-treatment observations, as detailed in Appendix C.3. The results remain consistent, further mitigating concerns about short pre-treatment periods for some treated communities. Third, our use of alternative estimators specifically designed to tackle staggered treatment timing and thus varying pre-treatment periods across communities, as detailed in Section 5.4.2, shows results consistent with the main findings. Lastly, the increasing popularity of LLMs and broader public adoption of AIGC during the observation period may confound the estimated effect of the AIGC ban. To assess this possibility, we conduct a placebo test that empirically examines whether unobserved confounding factors—such as rising AIGC awareness and use—could drive the results. As reported in Appendix C.4, the placebo treatment yields insignificant effects on all three outcomes, mitigating this concern.

Furthermore, we conduct additional robustness checks to address potential endogeneity concerns. First, to account for the possibility that differences in pre-existing AIGC activity and user concerns might confound the observed results, we compare the AIGC rate and discussions between the matched treated and control communities. The results in Appendix B.3 show that the matched groups exhibit comparable levels of AIGC usage and concerns, mitigating such issues. Second, to address the possibility that time-invariant unobservables and latent trends confound the estimates, we perform analyses based on look-ahead matching, using communities that banned AIGC at a later stage as the control group. The detailed procedure and the results are provided in Appendix C.5, and our findings remain robust. Third, treated communities may imitate one another's ban decision, which could confound the results. While our previous analyses already mitigate imitation concerns related to observed and unobserved community characteristics, we further validate the results by controlling for exposures to other AIGC bans. The results reported in Appendix C.6 remain robust. In summary, these comprehensive robustness checks yield consistent results and further strengthen the causal interpretation of the findings.

## 6. Exploration of Underlying Mechanisms

According to the theorization in Section 3, we conduct additional analyses to uncover the underlying mechanism. We begin by examining how AIGC reliability and social interactivity within a community shape user responses to the ban, and then analyze norm compliance and adaptation in knowledge seeking and knowledge contribution post-ban.

### 6.1 AIGC Reliability

As theorized in Section 3, information reliability is a central concern of AIGC applications and a key factor influencing how knowledge seekers and contributors respond to the AIGC ban (OpenAI 2023; Phang et al. 2009). In this subsection, we empirically test the role of information reliability by examining the heterogeneous effects of the AIGC ban on communities with high versus low AIGC reliability. If content reliability drives the observed changes in knowledge seeking and contribution efficiency, we expect communities with differing levels of AIGC reliability to respond differently to the ban.

Specifically, regarding knowledge seeking, the ban is expected to motivate more questions in communities with low AIGC reliability. In these communities, the ban increases the relative availability of reliable, human-created answers and signals an improvement in content quality. Additionally, online Q&A communities may gain a comparative advantage by addressing questions that LLMs cannot answer well, further motivating knowledge seeking. By contrast, the ban likely has little effect on knowledge seekers in communities with already reliable AIGC because the promotion of human-created answers provides minimal additional improvement in content quality.

Regarding contribution efficiency, we expect a larger drop in communities with high AIGC reliability. In these communities, knowledge contributors can typically rely on LLMs to quickly generate satisfactory answers when AIGC is allowed. The ban forces contributors to engage in a more effortful and time-consuming process to produce answers, leading to reduced answer efficiency. In contrast, in communities with low AIGC reliability, contributors already rely primarily on human effort to provide satisfactory answers, both before and after the ban. Consequently, the AIGC ban is likely to have little impact on their answer efficiency.

We empirically test the hypothesized effects by constructing a measure of AIGC reliability in each community. Constructing such a measure is challenging due to the diversity of topics and the absence of a well-established computational approach to calculate the AIGC reliability at the topic level. Following prior work (OpenAI 2023), we obtain an AIGC reliability score through human annotation using the following steps. First, we randomly sampled up to 10 questions with accepted answers from each Stack Exchange community, treating the accepted answers as correct responses. To ensure high-quality, human-created reference answers, we collect all sampled questions posted before the release of ChatGPT and remove any questions whose accepted answers were likely AI-generated. In total, 1,612 questions were sampled across all communities.[9] Second, we obtained ChatGPT's responses to these sampled questions. Notably, ChatGPT's training data only extends through September 2021[10], while our data spans from November 2022 to April 2023. This non-overlap allows us to address the concern that the LLM may have already 'known' the correct answers in advance, and it reflects the realistic scenario in which users use LLMs to answer new questions.

Using the accepted answers as references, two human annotators with graduate-level expertise rated the reliability of ChatGPT-generated answers for each question on a scale from 1 (lowest reliability) to 5 (highest reliability). Given the wide range of topics on Stack Exchange, annotators were permitted to consult external resources such as search engines to inform their evaluations. The average reliability score across all questions was 3.375 out of 5.0. The Cohen's kappa coefficient between the two annotators was 0.667, indicating substantial inter-rater agreement (Bao and Datta 2014).

Using the average AIGC reliability score in a community, we classify communities into high or low AIGC reliability groups based on the median score. We then replicate the analyses for each group to examine whether the effects of the AIGC ban differ by reliability level. As shown in Table 6, we find that the increase in question volume is concentrated in communities with low AIGC reliability, whereas the decrease in answer efficiency is observed only in communities with high AIGC reliability. These findings are consistent with our expectations and confirm that AIGC reliability contributes to the divergent responses in knowledge seeking and contribution efficiency following the ban.

---

[9] A few communities have fewer than ten questions with accepted answers due to their small size. In this case, we used all the questions from these communities.

[10] https://platform.openai.com/docs/models/gpt-3-5-turbo (last accessed on February 28, 2026).

In addition to the permutation test reported in Table 6, we conduct further analyses in Appendix D.2 to compare the estimated changes in communities with high and low AIGC reliability. The results remain consistent. We also complement this analysis with an examination of heterogeneity in common versus niche questions, detailed in Appendix D.4, which further supports the role of AIGC reliability.

**Table 6. Communities with High vs. Low AIGC Reliability**

| | (1) Log(NumQuestion) | (2) Log(NumAnswerPerQ) | (3) WithAcceptedAsw |
|---|---|---|---|
| Panel A: Communities with High AIGC Reliability | | | |
| $AIGC_Ban_{it}$ | 0.054 (0.064) | 0.012 (0.039) | -0.051*** (0.019) |
| Constant | 2.878*** (0.009) | 0.687*** (0.006) | 0.151*** (0.003) |
| No. Observations | 1,026 | 984 | 984 |
| Panel B: Communities with Low AIGC Reliability | | | |
| $AIGC_Ban_{it}$ | 0.224*** (0.055) | -0.041 (0.026) | -0.005 (0.024) |
| Constant | 2.648*** (0.004) | 0.671*** (0.002) | 0.185*** (0.002) |
| No. Observations | 874 | 847 | 847 |
| *p-value* of the permutation test (500 times bootstrap) | 0.000*** | 0.008*** | 0.01** |
| Community Fixed Effects | Y | Y | Y |
| Week Fixed Effects | Y | Y | Y |

Note: Robust standard errors are clustered at the community level and reported in parentheses; * $p<0.1$; ** $p<0.05$; *** $p<0.01$.

### 6.2 Social Interactivity

We next examine the role of social interactivity. As theorized in Section 3, the AIGC ban signals increased human involvement and, consequently, enhanced social interactivity in the communities. If social interactivity drives community response to the ban, communities with high versus low expectations of social interaction are expected to react differently.

Specifically, the increase in knowledge seeking is likely to be more pronounced in communities with strong expectations and established norms of social interactivity. In these communities, the promotion of human-created answers following the AIGC ban better satisfies the social needs of community members. Moreover, the ban may enhance the comparative advantage of online communities by supporting authentic social interactions, differentiating them from alternative LLM-based tools. This differentiation attracts and motivates knowledge seekers who value social interactivity, leading to more question asking. In contrast, communities that place less emphasis on social interactivity are relatively indifferent between human-created answers and AIGC, and knowledge-seeking behavior in these communities is likely to remain unchanged following the ban.

For knowledge contributors, AIGC is more likely to be readily used by contributors in communities with less emphasis on social interactivity. In these communities, the ban forces contributors to invest additional effort and time to produce satisfactory answers, thereby reducing answer efficiency. In contrast, in communities with a richer social fabric and stronger expectations of interpersonal engagement, contributors are likely less affected by the ban because substantial human input is already required and remains as the primary driver for producing satisfactory answers both before and after the ban.

In our empirical analyses, we measure the expected level of social interactivity in a community based on the question content. Specifically, we use the socialness scores for English words developed by Diveica et al. (2023), which indicate the degree to which a word is socially relevant—a higher score corresponds to stronger social relevance. Using questions posted before the AIGC ban, we calculate the average socialness score for each question, with higher score reflecting a greater emphasis on authentic social interactions and human-provided answers. We then classify communities as high- or low-social interactivity based on the median value of these scores and estimate post-ban changes separately for each group, with the results reported in Table 7.

Several interesting findings are observed from Table 7. Regarding knowledge seeking, as expected, we observe an increase in the number of questions posted in communities with high social interactivity, whereas question volume in communities with lower social interactivity remains unchanged following the AIGC ban. These findings align with Burtch et al. (2024), who document a significant decrease in knowledge seeking after ChatGPT's release in communities with a weaker social fabric, but not in those with a stronger social fabric. Both Burtch et al. (2024) and our study share the theoretical rationale that users vary in their social needs and select tools that best satisfy those needs. Following ChatGPT's release, users in communities with less emphasis on social interactivity tend to be indifferent between online communities and ChatGPT, making it easier for them to shift knowledge seeking to ChatGPT for faster answers, as noted in Burtch et al. (2024). In the case of the AIGC ban, users in communities with stronger expectations of social interactivity perceive greater benefits from engaging in online Q&A communities after the ban, which drives increased question asking in these communities, as reflected in our analyses.

**Table 7. Communities with High vs. Low Social Interactivity**

| | (1)<br>Log(NumQuestion) | (2)<br>Log(NumAnswerPerQ) | (3)<br>WithAcceptedAsw |
|---|---|---|---|
| Panel A: Communities with high social interactivity | | | |
| $AIGC_Ban_{it}$ | $0.113^{**}$ | -0.034 | -0.016 |
| | (0.052) | (0.028) | (0.014) |
| Constant | $3.851^{***}$ | $0.744^{***}$ | $0.206^{***}$ |
| | (0.007) | (0.004) | (0.002) |
| No. Observations | 950 | 912 | 912 |
| Panel B: Communities with low social interactivity | | | |
| $AIGC_Ban_{it}$ | 0.138 | 0.041 | $-0.069^{**}$ |
| | (0.099) | (0.059) | (0.03) |
| Constant | $1.689^{***}$ | $0.616^{***}$ | $0.129^{***}$ |
| | (0.009) | (0.005) | (0.003) |
| No. Observations | 950 | 919 | 919 |
| *p-value* of the permutation test (500 times bootstrap) | 0.402 | $0.056^{*}$ | $0.058^{*}$ |
| Community Fixed Effects | Y | Y | Y |
| Week Fixed Effects | Y | Y | Y |

Note: Robust standard errors are clustered at the community level and reported in parentheses; * $p<0.1$; ** $p<0.05$; *** $p<0.01$;

Regarding knowledge contribution, the number of answers per question remains stable across communities regardless of their level of social interactivity. However, consistent with our expectations, we observe a decrease in answer efficiency in communities with low expectations of social interactions following the AIGC ban, while communities with high social interactivity show no significant changes. We further validate these findings using an alternative, activity(comment)-based measure of expected social interactivity in a community. We also apply alternative analyses in addition to the permutation tests to compare the changes in communities with high and low social interactivity. Appendix D.3 details the procedure and results of these analyses. Our results remain largely consistent.

**6.3 Additional Mechanism Exploration**

Additionally, as theorized in Section 3.4, the AIGC ban signals that human involvement is the expected norm in the community. Knowledge seekers and contributors may adapt their posted content to comply with this norm, and such adaptation could contribute to the decrease in contribution efficiency observed in the main analyses. In this subsection, we conduct several analyses to examine norm compliance and adaptation following the ban.

We begin by examining shifts in the types of questions posted by knowledge seekers. The AIGC ban emphasizes the increased prevalence of human-created answers and distinguishes online Q&A communities from alternative knowledge sources, such as LLMs. As a result, the ban may amplify knowledge seekers' expectations for human-generated content and motivate them to ask questions that are better suited for humans to answer, such as questions that are more reliably answered by humans or that require greater social awareness and nuanced understanding.

To examine these changes, we construct two sets of variables that capture question askers' expectations regarding reliable information and social input from human contributors. First, with respect to the informational aspect, we create two variables: (i) question length, which captures the information richness and potential complexity of the inquiry; and (ii) subjectivity, which reflects the extent to which a question is framed around personal opinions or subjective concerns. LLMs tend to provide less reliable answers to longer, more complex, or highly subjective questions (Saxena et al. 2024; Wang et al. 2025), making human-generated answers more desirable for such questions.

Second, with respect to the social aspect, we construct a variable to capture the 'socialness' of a question. To this end, we employ socialness scores for English words developed by Diveica et al. (2023). A high score reflects stronger social relevance. We calculate the average socialness score of each question, where a high score signals greater emphasis on and preference for authentic human answers.

Using the above variables, we estimate changes in posted questions following the AIGC ban. Columns (1)-(3) of Table 8 show that, on average, questions in a community become longer, more subjective, and exhibit higher socialness post-ban, suggesting that knowledge seekers ask questions that are more suitable for humans to answer. Notably, this effect is only evident in non-STEM communities. These content changes complement our findings on the moderating roles of AIGC reliability and social interactivity. Together, the evidence suggests that the AIGC ban amplifies knowledge seekers' expectations for human-generated content in online Q&A communities. Knowledge seekers adapt their questions accordingly to better satisfy their information and social needs and to align with the anticipated increase in human contribution.

**Table 8. Impacts of the AIGC Ban on the Content of Posted Questions and Answers**

| | Questions | | | Answers | | |
|---|---|---|---|---|---|---|
| | (1) Length | (2) Subjectivity | (3) Socialness | (4) Length | (5) Subjectivity | (6) Socialness |
| Panel A: All Communities | | | | | | |
| $AIGC_Ban_{it}$ | $0.156^{***}$ | $0.12^{***}$ | $0.151^{**}$ | 0.057 | 0.036 | 0.043 |
| | (0.051) | (0.043) | (0.059) | (0.063) | (0.053) | (0.071) |
| No. Observations | 1,831 | 1,831 | 1,831 | 1,831 | 1,831 | 1,831 |
| Panel B: STEM Communities | | | | | | |
| $AIGC_Ban_{it}$ | 0.06 | 0.059 | 0.046 | -0.07 | -0.059 | -0.072 |
| | (0.08) | (0.064) | (0.082) | (0.083) | (0.071) | (0.100) |
| No. Observations | 1,042 | 1,042 | 1,042 | 1,042 | 1,042 | 1,042 |
| Panel C: Non-STEM Communities | | | | | | |
| $AIGC_Ban_{it}$ | $0.28^{***}$ | $0.224^{***}$ | $0.29^{***}$ | $0.210^{***}$ | $0.176^{***}$ | $0.181^{***}$ |
| | (0.05) | (0.05) | (0.072) | (0.058) | (0.057) | (0.064) |
| No. Observations | 789 | 789 | 789 | 789 | 789 | 789 |
| Community Fixed Effects | Y | Y | Y | Y | Y | Y |
| Week Fixed Effects | Y | Y | Y | Y | Y | Y |

Note: Robust standard errors are clustered at the community level and reported in parentheses; * $p$<0.1; ** $p$<0.05; *** $p$<0.01.

We also examine content changes in answers posted by knowledge contributors. Given that the AIGC ban emphasizes human effort in answer creation, contributors are expected to invest more time and effort in producing responses. In addition to aligning with these expectations, contributors may also adapt their answers in response to changes in the content of questions. To empirically test knowledge contributors' behavioral adaptation, we use the same set of variables employed in the question content analysis and report the results in Columns (4)-(6) of Table 8. Across all questions, answer characteristics remain unchanged. However, when focusing on non-STEM communities in which answer efficiency declined, we find that answers become longer, more subjective, and exhibit higher socialness, suggesting increased human efforts and time spent on answer creation post-ban.

We further analyze changes in the characteristics of question askers and answer providers, with detailed analyses reported in Appendix D.6. Our results indicate that questions are increasingly asked by more experienced users following the ban. Similarly, more seasoned users provided answers post-ban. These patterns align with the observed content changes: the amplified expectation of human involvement in questions is driven by more experienced askers, who are likely less satisfied with AIGC compared to less experienced users (Burtch et al. 2024). Correspondingly, answers are provided by more experienced users, who are more likely to value human input and capable of tackling more complex and nuanced questions (Li and Kim 2025).

The analyses of content changes in questions and answers collectively suggest that knowledge seekers and contributors engage in norm compliance, adapting to the increased expectation of human involvement following the ban. These results also indicate that such behavior adaptation is an additional driver of the post-ban decline in contribution efficiency, as answer providers must invest more effort and time to produce satisfactory answers to more complex and socially nuanced questions.

The analyses also provide insights into the heterogeneity across community types. Content changes in questions and answers are consistently observed in non-STEM communities but not in STEM ones. This pattern aligns with the differing nature of these community types: in non-STEM communities where knowledge seeking often emphasizes more context-aware expertise, authentic human experience, and emotional intelligence, the AIGC ban appears to amplify user expectations for such content, as reflected in the observed shifts in posted content. In contrast, STEM communities, where questions are typically more factual and place less emphasis on social interactions, show little change in posted content after the ban.

## 7. Discussion and Conclusion

With the increasing adoption of AIGC across digital platforms, some online Q&A communities, among the most visited platforms, implemented the AIGC ban in response to growing concerns about AIGC. Our study leverages the staggered AIGC ban decisions across several Stack Exchange communities to examine the consequences of such bans. The findings suggest that banning AIGC has a double-edged effect: while it increases question asking, it simultaneously reduces answer efficiency. Notably, these effects are observed only in non-STEM communities.

Our analyses further show that AIGC reliability and a community's emphasis on social interactivity play important roles in shaping the behavior of question askers and answer providers following the ban. Increased question volume occurs in communities where human answers are perceived to be more reliable and better satisfy social needs. Conversely, reduced answer efficiency is observed in communities with high AIGC reliability but low emphasis on social interactivity, where AIGC was likely used with minimal human effort. Additionally, both question askers and answer providers adapt their questions and answers to align with the community's heightened expectations for human involvement post-ban.

### 7.1 Theoretical Implications

Our work contributes to three streams of research. First, we extend the growing body of literature on the dynamics between AIGC and online Q&A communities by underscoring the trade-offs associated with banning AIGC in several aspects. On one hand, banning AIGC would lead to increased knowledge seeking, reversing some of the effects of LLM release (Burtch et al. 2024; Quinn and Gutt 2025). On the other hand, the AIGC ban reduces contribution efficiency in online communities, a downside that has received less attention in the literature (Wang and Zheng 2024). By leveraging the heterogeneity in our dataset, we show that these trade-offs only exist in non-STEM communities, where LLM-generated responses appear adequate, but human-contributed answers are more expected. Additionally, our work extends this line of work by empirically demonstrating that AIGC reliability and social interactivity are two critical factors influencing user behavior shifts in these communities. Overall, our study offers valuable insights for future research on the interplay between human-created online knowledge systems and AIGC, particularly on examination of platform performance, community heterogeneity, and mechanisms.

Second, our work extends the literature on machine substitution and augmentation (Autor and Dorn 2013; Acemoglu and Restrepo 2018; Dixon et al. 2021) by underscoring the value of a socio-technical perspective in understanding the interplay between humans and machines. The extant literature primarily considers that the extent to which machines substitute for or augment humans depends on task characteristics, such as whether tasks are routine, codified, and objective. In contrast, a socio-technical perspective treats online communities as socio-technical systems, offering a distinct lens to examine the interaction between humans and machines that involves both social and technical mechanisms. Particularly in the context of online Q&A communities, our work shows that AIGC tends to substitute for human-generated answers in topics where it is reliable and in less socially oriented communities. However, AIGC struggles to replace human contributions in topics with low reliability or in interaction-oriented environments.

Lastly, our work contributes to the literature on platform competition and differentiation (Cennamo and Santalo 2013; Rietveld and Schilling 2021) by highlighting that repositioning along the technical and social dimensions as a critical differentiation strategy for human-centered platforms amid rising generative AI adoption and competition. Platforms gain comparative advantages by repositioning their content offerings and platform experiences toward areas where human involvement provides distinct value. Most importantly, because generative AI functions as both a substitute and cost-reducing technology that offers distinct value to different platform participants, our study provides empirical evidence of a critical trade-off between engagement and efficiency following platform repositioning. These findings shed light on the growing body of research examining competition between traditionally human-created platforms and external generative AI systems.

### 7.2 Practical Implications

Our research offers important practical implications. First, our results highlight that banning AIGC in communities that traditionally rely on human contribution —a common strategy for online communities to address the growing impact of AIGC—cannot fully reverse the effects of LLM release. Platform managers should recognize that LLMs affect answerers and askers differently, particularly in communities focused on non-STEM topics. Accordingly, platform administrators should identify which aspect of the community they aim to strengthen and implement corresponding policies aligned with that goal. If the objective is to encourage more questions, banning AIGC can be effective. Conversely, if the community's growth is constrained by answer inefficiencies, relaxing restrictions on AIGC can improve overall answer efficiency and community sustainability.

Moreover, when using an AIGC ban to differentiate online communities that emphasize authentic human-created content, platform managers must consider changes in posted content and user composition following the ban. In communities where human involvement becomes more effortful following the ban, complementary policies are needed to safeguard user experience. For example, communities can offer greater intrinsic or extrinsic rewards to incentivize answer providers to tackle complex questions with higher social nuance. Additionally, managers should actively promote questions to domain experts, such as by sending invitations or making targeted recommendations, to ensure the increasing informational and social needs of question askers are met.

Additionally, our study highlights the potential for a mixed AIGC policy in online Q&A communities to improve knowledge exchange. Such a policy would help users distinguish between questions suitable for AIGC and those better addressed by humans, based on AIGC reliability and the degree of social interactions emphasized by the question. For example, platform managers could label questions with low complexity or minimal social nuance as "open for AIGC," facilitating efficient answer contribution since LLMs are more likely to provide reliable responses to these queries. Conversely, questions with high complexity or strong social nuance could be labeled 'human answer only' to prevent the spread of potentially unreliable AI-generated answers and encourage high-quality human contributions.

### 7.3 Limitations and Future Work

Our work has several limitations for future research. First, we rely on archival data from Stack Exchange, one of the world's largest Q&A communities. While informative, other digital platforms may adopt different approaches to address the challenges posed by generative AI. Future researchers could extend this work by examining the effects of alternative platform policies and content moderation strategies on user behavior. Moreover, given the difficulty of fully eliminating endogeneity concerns in studies using archival datasets, researchers could further validate our findings using alternative identification strategies.

Second, despite our efforts to use both content-based and activity-based measures to proxy a community's expectation for social interactivity, these measures may not fully capture the intended construct. We encourage further empirical examination of the role of social interactivity when more appropriate measures become available.

Third, our study focuses on ChatGPT and its impact on knowledge exchange outcomes in Q&A communities. However, LLMs and, broadly, generative AI technologies, are advancing rapidly, so caution is warranted when generalizing our findings to future AI tools or other settings.

Finally, the integration of AIGC into user-generated content may vary. For example, LLMs can fully replace humans in generating answers or act as assistants, helping users brainstorm ideas or refine their responses. These differing modes of use may have varied effects on productivity, content quality, and user-perceived socialness in online Q&A communities (Chen and Chan 2024). Due to the observational nature of our data, we are unable to observe how AIGC was used in each instance. We encourage future research to explore this important direction.